\documentstyle[12pt,amssymb]{article}
\input epsf
\def\theequation{\arabic{section}.\arabic{equation}}
\newcounter{subequation}[equation]
\makeatletter
\@addtoreset{equation}{section}
\expandafter\let\expandafter\reset@font\csname reset@font\endcsname
  {\endeqnarray\stepcounter{equation}}
\makeatother

\normalbaselines
\begin{document}
\title{Chern--Simons Particles with Nonstandard Gravitational Interaction}

\author{J. Lukierski\\
Institute for Theoretical Physics, University of Wroc\l aw, \\
 pl. Maxa Borna 9, 50-204 Wroc\l aw, Poland\\ email: lukier@ift.uni.wroc.pl
\\
                     and\\
   Departamento de Fisica Teorica, Universidad de Valencia,\\
   Av. dr Moliner 50, E-46100 Burjasot (Valencia), Spain
         \\  \\
P.C. Stichel\\
An der Krebskuhle 21\\ D-33619 Bielefeld, Germany \\
e-mail:pstichel@gmx.de
\\ \\
W.J. Zakrzewski
\\
Department of Mathematical Sciences, University of Durham, \\
Durham DH1 3LE, UK \\
 e-mail: W.J.Zakrzewski@durham.ac.uk\\
and\\
Center for Theoretical Physics\\
Massachusetts Institute of Technology, 
Cambridge, 02139, USA}
\date{}
\maketitle
\begin{abstract}
The model of nonrelativistic particles coupled to nonstandard
(2+1)--gravity [1] is extended to include Abelian or
non-Abelian charges coupled to Chern--Simons gauge fields.
Equivalently, the model may be viewed as describing the (Abelian or
non-Abelian) anyonic dynamics of Chern--Simons  particles
coupled, in a reparametrization invariant way, to a translational
Chern--Simons action. The quantum two--body problem is described 
by a nonstandard Schr\"{o}dinger equation with a noninteger
angular momentum depending on energy as well as particle
charges. Some numerical results describing the
modification of the energy levels by these charges in the
confined regime are
presented. The modification involves a shift as well as splitting
of the levels.
\end{abstract}

\section{Introduction}

\textheight 8.5in \textwidth 6in

Particles in 2+1 dimensions ($D=2+1$)
carrying electric (Abelian) or isospin (non--Abelian) internal
charges coupled to Chern--Simons (CS) gauge fields, have been considered
in many applications (see e.g. [2,3]). These particles, in the
Abelian case describing anyons (see e.g. [4]) and for
non--Abelian couplings their generalizations (``non--Abelian
 anyons"), are characterized, respectively, by the Abelian
and
non--Abelian versions of braided fractional statistics (see e.g. [5,6]).

The aim of this paper is to supplement the dynamics of
nonrelativistic CS particles with nonstandard gravitational
interactions  described in [1]. The free field actions in our
model for the $D=2+1$ gravitational and gauge fields are described
by the CS Lagrangians. In particular, in the gravitational sector
described by dreibeins $E^{\underline{a}}_{\mu}$ $(\mu=0,1,2;\underline{ a}=
1,2)$, with
tangent space indices restricted to nonrelativistic
 $SO(2)$ space rotations, we use the following action proposed in [1]:
\begin{equation}
S^{\rm GR}_{0} = {1\over 2 \lambda} \int d^{3}x\, \epsilon^{\mu\nu\rho}
\, E^{\underline{a}}_{\mu}\, T^{\underline{{a}}}_{\nu\rho} + S_{\rm B} \, ,
\end{equation}
where $ T^{\underline{{a}}}_{\mu\nu}=\partial_{\mu}  E^{\underline{a}}_{\nu}- \partial_{\nu}
 E^{\underline{a}}_{\mu}$ describes the $D=2+1$ torsion field and $S_{\rm
B}$ are boundary terms  specified in [1,12]. The dreibeins
transform covariantly under local space translations (fixed
 time diffeomorphisms). Then the invariant free action for
nonrelativistic point particles\footnote{For reasons of
simplicity we give all particles the same mass $m=1$ in
appropriate units.} described by trajectories
$x^{i}_{\alpha}(t)$ ($i=1,2;$ $\alpha, = 1 \ldots N$) is  given,
in the first  order
 formalism, by  [1,12]
 \begin{equation}
 S^{({\rm N})}_{{\rm part},0} =
 \int dt\, \sum\limits^{N}_{\alpha=1}
 \left(  \xi^{\underline{a}}_{\alpha}
 \left(
  E^{\underline{a}}_{j,\alpha}
 \dot{x}^{j}_{\alpha} +
  E^{\underline{a}}_{0,\alpha}
  \right)
  - {1\over 2}
  \xi^{\underline{a}}_{\alpha} \xi^{\underline{a}}_{\alpha}
  \right)\, .
  \end{equation}

  In the gauge
sector we consider the known free CS actions:

i) Abelian case ($A_{\mu}$ - electromagnetic potential)

\begin{equation}
S^{\rm A}_{0} = {\kappa\over 4}
\int d^{3}x \,
\epsilon^{\mu\nu\rho}
\, A_{\mu} \, F_{\nu\rho} =
{\kappa\over 2}
\int d^{3}x \,
\epsilon^{\mu\nu\rho}
\, A_{\mu} \, \partial_{\nu}\, A_{\rho}\, ,
\end{equation}
where
$ F_{\mu\nu}=\partial_{\mu}  A_{\nu}- \partial_{\nu}
 A_{\mu}$

ii) Non-Abelian case ($A^{i}_{\mu} $ - isospin gauge field
potential; for simplicity we shall choose the internal symmetry
group $G=SU(2)$) (cp. [8]):
\begin{equation}
S^{\rm NA}_{0}     =
{\kappa\over 2}
\int d^{3}x \,
\epsilon^{\mu\nu\rho}
\,\left(
 A^{i}_{\mu} \, \partial_{\nu}\, A^{i}_{\rho}
 + { 1\over 3}
 \epsilon_{ijk}\,
 A^{i}_{\mu}  A^{j}_{\nu}  A^{k}_{\rho}\right) \, .
\end{equation}

In the Abelian case the ``charge space" is trivial, described by
a constant numerical parameter.
In the non--Abelian case the internal degrees of freedom of CS
particles should be explicitly taken into consideration by
extending the space--time geometry (see e.g. [9--10]). The
 non--Abelian charge space coordinates $Q^{i}(t)$ carrying the
adjoint representation of internal symmetry group $G$ (in our
case $i=1,2,3$ and $Q^{i}$ is the $SO(3)$ isovector), after
quantization ($Q\sp{i}\rightarrow \hat Q\sp{i}$), constitute
the quantum mechanical analog of current algebra coordinates,
with Lie--algebraic equal time commutation relations
\begin{equation}
\left[ \hat Q^{i}(t),\hat  Q^{j}(t) \right]
= i \, \epsilon^{ijk} \, \hat Q\sp{k}(t)\, .
\end{equation}
It is convenient to put the coordinates $Q^{a}$ on the sphere
$S^{2}$ of radius $J$:
\begin{equation}
Q^{i}\, Q^{i} = J^{2}\, ,
\end{equation}
which describes an adjoint symplectic  orbit
 of $SU(2)$ with the following Kirillov symplectic two form [9]
 \begin{equation}
 \Omega = {1 \over 2J^{2}}
 \epsilon^{ijk}\, Q^{i} \, d Q^{j}\, d Q^{k} \, .
 \end{equation}

 After quantization the relation (1.6) defines the Casimir of
$SU(2)\simeq SO(3)$ algebra (1.5)  which implies the quantization of the
radius $J$ by integers and half--integers. Using Darboux
variables one can derive from (1.7) the free action for charge
coordinates $Q^{i}$ (see Sect. 3). We would like to recall here
that the dynamics of free particles on space--time $\times
S^{2}$ manifold was first derived in the Kaluza--Klein
framework [11] leading to $Q^{i}$ which satisfy
 the Wong equations [10].

This paper can be regarded as the extension of our results in
[1,12], where we have considered the interaction of $D=2$
nonrelativistic particles and of the gravitational field $E^{\underline{a}
}_{\mu}$
governed by the free action (1.1). In [12] (see Sect. 8) we have considered
also the interaction with a constant $D=2$ magnetic field.
In this paper we consider the additional dynamical Abelian and non--Abelian
CS gauge fields. Interestingly enough the coupling to dynamical
gauge fields appears simpler than  the interaction
with fixed external gauge potentials. This simplicity  follows from
the main property of the CS interactions in $D=2+1$,   the local
field -- current identity, which permits us to solve algebraically
the fields in terms of the sources (see e.g. [7,8]). In
consequence, we obtain the solvability of the $D=2+1$ two--body
problem with gravitational and CS gauge interactions. It should
be mentioned here that an analogous problem for $D=1+1$ [13,14] can
be also solved; however, the field--current identity looses its
local character (see [15]).

The plan of our paper is as follows:

In Sect. 2 we extend the model given in [1] by including the  coupling to
the $D=2+1$ CS electrodynamics (see (1.3)) and present the classical
dynamics.

In Sect. 3 we
 consider the extension of the notion of point particles in
space--time to non--Abelian CS particles ([8--11] with
space--time points supplemented by internal charge coordinates
$Q^{i}$, constrained by (1.6).
The non--Abelian gauge sector is described by the CS action
(1.4). We find that, after quantization, the energy levels of the
  two--body problem  depend on the eigenvalues of the following
isospin-like  operator
\begin{equation}
{\hat \Omega\over 2}\,:=\,\hat Q^{i}_{1}\,\hat  Q^{i}_{2} =
{1\over 2} \left(\hat J^{2}_{12} -\hat  J^{2}_{1} -\hat J^{2}_{2}
\right)\, , 
\end{equation}
where
$\hat J^{2}_{12}=\left(\hat Q^{i}_{1} + \hat Q^{i}_{2}\right)^{2}$,
$\hat J^{2}_{1}=\left(\hat Q^{i}_{1} \right)^{2}$,
$\hat J^{2}_{2}=\left(\hat Q^{i}_{2} \right)^{2}$. Thus, if the eigenvalues
of the individual particles are $j_1(j_1+1)$ and $j_2(j_2+1)$
respectively, the eigenvalues of $\hat J_{12}\sp2$, are given by $j_{12}(j_{12}+1)$
where $j_{12}$ lies between $\vert j_1+j_2\vert$ and $\vert j_1-j_2\vert$.
In both cases of electric and isospin interactions, in Sect. 2
and Sect. 3 we  present the modification of the classical
results given in [1,12] which did not include  the gauge interactions. In
Section 4 we describe two--body quantum mechanics and present
numerical results for the corresponding modification of the
energy spectra in the confined regime for both the Abelian and
the non-Abelian case.


Sect. 5  presents some outlook. 

\section{Classical Dynamics for the $D=2$ Abelian Case}

We  consider the following action of $N$ nonrelativistic
 charged
particles interacting with dreibein fields $E^{a}_{\mu}$ and
an electromagnetic field $A_{\mu}$:

\begin{equation}
S^{({\rm N})}_{\rm part} = S^{({\rm N})}_{{\rm part},0}
+ \int dt
\sum\limits^{N}_{\alpha=1}
e_{\alpha}
\left(
A_{j,\alpha} \dot{x}^{\ j}_{\alpha} +
A_{0,\alpha} \right)\, ,
\end{equation}
where $ e_{\alpha}$    is the electric charge of the
$\alpha$--th particle. Under the assumption that the
fields $A_{\mu}(\vec{x},t)$ transform covariantly under fixed
 time
diffeomorphism  $S^{({\rm N})}_{\rm part}$ is an invariant entity.

The full action  is given now by

\begin{equation}
S^{(\rm N)} = S^{\rm GR}_{0}
+ S^{\rm A}_{0}+
 S^{(\rm N)}_{\rm part}\, .
\end{equation}
The equation of motion (EOM), the Gauss 
constraint for the dreibeins $E^{\underline{a}}_{\mu}$ derived from (2.2) and
their solution are described in [1,12]. We have
\begin{equation}
E^{\underline{a}}_{\mu}(\vec{x},t) = - {\lambda \over 4\pi}
\partial_{\mu} \sum\limits_{\alpha} \,  \xi^{\underline{a}}_{\alpha}
 \, \phi
 \left(\vec{x} - \vec{x}_{\alpha}\right) +
 E^{as,\underline{a}}_{\mu} \, ,
 \end{equation}
 with

\begin{equation}
 E^{as,\underline{a}}_{i}  =\delta^{a}_{i}\, ,
 \end{equation}

\begin{equation}
 E^{as,\underline{a}}_{0} =  - v^{\underline{a}} (t)\, ,
 \end{equation}
where the singular gauge function $\phi$ is defined by
\begin{equation}
\phi(\vec{x}): = arc \, tan  {x_{2}\over x_{1}}\, ,
\end{equation}
and regularized in  such a way that $\partial_{k}\phi(\vec{x})$ vanishes for
$\vec{x}\to {0}$.

Variation of $S$ with respect to $v^{\underline{a}}(t)$ leads to
the constraint
\begin{equation}
\sum\limits_{\alpha} \xi^{\underline{a}}_{\alpha} = 0\, ,
\end{equation}
and therefore to the  vanishing of the
total momentum of the $N$--particle system [1,12].

The choice of gauge (2.4--5) for the
dreibeins breaks asymptotically the invariance with  respect to
local space translations leaving, as residual symmetry [1,12],
only translations  local in time,
and rigid rotations.

The EOM and the Gauss 
 constraint for the $A_{\mu}$, and their solutions, are all well known
(cp. [7]). We have
\begin{equation}
A_{\mu}(\vec{x},t) =
 - {1 \over 2\pi \kappa}
\partial_{\mu} \sum\limits_{\alpha} \, e_{\alpha}
\,
 \phi
 \left(\vec{x} - \vec{x}_{\alpha}\right) \, .
 \end{equation}
Let us consider now the two--body case i.e. $N=2$ in detail.

Applying the Legendre
 transformation to the Lagrangian (2.2) and using the relevant
constraints (Gauss 
 and (2.7)) we obtain for the two--body Hamiltonian $H$
 describing relative motion
 \begin{equation}
 H= \xi_{i} \, \xi_{i}\, ,
 \end{equation}
 where we have defined
\renewcommand\theequation{\thesection.9\alph{equation}}
\setcounter{equation}{0}
 \begin{equation}
 \vec{\xi} : = {1\over 2} \left(  \vec{\xi}_{1}
  -   \vec{\xi}_{2} \right)\, .
  \end{equation}

  Denoting the canonical particle momenta by $\vec{p}_{\alpha}$
  and defining
  \begin{eqnarray}
  \vec{p} &:= & {1\over 2} \left(\vec{p}_{1} -\vec{p}_{2} \right)
  \\ 
    \vec{x} &:= & {1\over 2} \left(\vec{x}_{1} -\vec{x}_{2} \right)
    \end{eqnarray}
    we obtain from (1.2), (2.1), (2.3--5) and (2.8--9) the relation
\renewcommand\theequation{\thesection.\arabic{equation}}
\setcounter{equation}{9}
    \begin{equation}
    \xi_{i} = p_{i} - {\lambda \over 4\pi}
    \partial_{i} \, \phi(\vec{x})
    \left( H- {2 \over \lambda \kappa} e_{1} \, e_{2}
    \right)\, .
    \end{equation}
    Squaring it and using again (2.9) we obtain
\begin{equation}
H = p^{2} - {l^{2}\over r^{2}} +   {\overline{l}^{2}\over
r^{2}}\, ,
\end{equation}
where the angular momentum for the relative motion $l$
($l:=\vec{x}\wedge  \vec{p}$) is, according to (2.10), given by
    \begin{equation}
    l = \overline{l} +
     {\lambda \over 4\pi}
    \left( H- {2 \over \lambda \kappa} e_{1} \, e_{2}
    \right)\, ,
    \end{equation}
with
 \begin{equation}
\overline{l}:=\vec{x}\wedge  \vec{\xi}\, .
\end{equation}

Note that (2.11) has the same form in case of absence of the
Abelian gauge fields [1,12], only the relation (2.12) gets an
additional term.

By applying an inverse
 Legendre transformation to (2.9) and
using (2.10) we obtain the well known result that the coupling
to the Abelian gauge fields leads only to the addition of a total
time derivative to the two--particle Lagrangian 
\begin{equation}
L = L_{0} - {e_{1} e_{2} \over 2\pi \kappa}
 \, {d\over dt} \, \phi(\vec{x})\, .
 \end{equation}
 Therefore the classical EOM are unchanged in comparison      
with $e=0$ case [1,12].
Due to the singular nature of ${d \over dt} \phi$ this holds for
noncoinciding particle
positions, i.e. for $\vec{x} \neq
\vec{0}$ only.

In particular, we conclude from [1,12] that:

i)  \begin{equation}\dot{\xi}_{i} = 0 \, ,
\end{equation}

ii)

\begin{equation}
\dot{x}_{i} = {2  \xi_{i} \over 1 + { \lambda \overline{l}
\over 2\pi r^{2}\, .
}}
\end{equation}
leading to  $\overline{l}$  being a
conserved quantity and to the geometric bag formation in the case of
$\lambda \overline{l} < 0$
for 
\begin{equation}
r <r_{0}:= \left(
{\lambda \overline{l} \over 2 \pi} \right)^{1/2}
\end{equation}

But, as shown in Section 4, the additional term in (2.12) leads,
in the quantum case, via (2.11) to a modification of the energy
levels in the confined regime.

\section{Classical Dynamics for the $D=2$ Non--Abelian Case}

We now consider the interaction of $N$ nonrelativistic
particles carrying $SU(2)$--charges  $Q^{a}$ ($a=1,2,3$) with
dreibein fields ${E} ^{\underline{a}}_{\mu}$
and $SU(2)$--gauge fields $A^{a}_{\mu}$. The corresponding
particle action $S^{({\rm N})}_{\rm part}$ is given by
\begin{equation}
S^{({\rm N})}_{\rm part} = S^{({\rm N})}_{{\rm part},0} +
\int dt \, \sum\limits^{N}_{\alpha=1}
\, Q^{a}_{\alpha}
\left( A^{a}_{j,\alpha} \dot{x}^{j}_{\alpha} +
A^{a}_{0,\alpha} \right) +
S^{({\rm N})}_{SU(2)}\, ,
\end{equation}
where  $S^{({\rm N})}_{SU(2)}$ is the action which   is  given by
the symplectic form (1.7) [9].
Choosing on the   sphere $S^{2}$ the spherical coordinates  
 one gets
\begin{equation}
S^{({\rm N})}_{SU(2)} := \int dt
 \sum\limits^{N}_{\alpha=1} \cos \theta_{\alpha}(t)
 \dot{\phi}_{\alpha} (t)\, ,
 \end{equation}
 with $\theta, \phi$ being the angles on the sphere $S^{2}$.

The total action is now given by (see (1.1), (1.4) and (3.1-2)) 
\begin{equation}
S^{({\rm N})} =
S^{({\rm GR})}_{0} + S^{({\rm NA})}_{0} + S^{({\rm N})}_{\rm
part}\, .
\end{equation}
As again, the gauge and gravitational degrees of freedom  are not
coupled directly and, as in the Abelian case, the dreibeins are described by (2.3--5).

The Euler--Lagrange equation for the $SU(2)$--gauge fields
$A^{a}_{\mu}$ are given by [16]:

i) the Gauss constraint
\begin{equation}
F^{a}_{ij} (\vec{x},t) =
- {1\over \kappa} \, \epsilon_{ij} \,
\sum\limits_{\alpha} \, Q^{a}_{\alpha} \,
\delta(\vec{x} - \vec{x}_{a} )\, ,
\end{equation}
and

ii) the EOM
\begin{equation}
F^{a}_{i0} (\vec{x},t) =
 {1\over \kappa} \, \epsilon_{ij} \,
\sum\limits_{\alpha} \, Q^{a}_{\alpha} \,
\dot{x}^{j}_{\alpha}
\delta(\vec{x} - \vec{x}_{a} )\, ,
\end{equation}
where $F^{a}_{\mu\nu}$ is the $SU(2)$--field strength
\begin{equation}
F^{a}_{\mu\nu} : =
\partial_{\mu} \, A^{a}_{\nu} -
\partial_{\nu} \, A^{a}_{\mu} +
\epsilon_{abc}
\,  A^{b}_{\mu}\,  A^{c}_{\nu}\, .
\end{equation}
Usually the nonlinear
 Gauss 
constraint (3.4) is solved in the axial gauge $ A^{a}_{1} \equiv
0$, but then we would loose the rotational covariance.

For the  derivation of the   effective two--body dynamics we  need only
the $A^{a}_{i}(\vec{x},t)$ at the particle positions
 $\vec{x}_{1,2}$. Fortunately, at these positions, the
solution of (3.4) may be obtained explicitly [16]   in  the  following form:
\renewcommand\theequation{\thesection.7\alph{equation}}
\setcounter{equation}{0}
\begin{eqnarray}
A^{a}_{i,1} &=& - {1\over 2\pi \kappa} \, Q^{a}_{2} \,
\partial_{i} \, \phi \left(\vec{x}_{1} - \vec{x}_{2} \right)
\\
A^{a}_{i,2} & =&  {1\over 2\pi \kappa} \, Q^{a}_{1} \,
\partial_{i} \, \phi \left(\vec{x}_{1} - \vec{x}_{2} \right)
\end{eqnarray}
In order to make the arguments given in [16] rigorous
 we introduce a gauge field $\widetilde{A}\ ^{a}_{i}$
 \begin{equation}
 \widetilde{A}\ ^{a}_{i}(\vec{x},t):=
  - {1\over 2\pi \kappa} \, \sum\limits^{2}_{\alpha=1}
   Q^{a}_{\alpha} \,
\partial_{i} \, \phi \left(\vec{x} - \vec{x}_{\alpha} \right)\, ,
 \end{equation}
 solving the linearized Gauss constraint (3.4).

\renewcommand\theequation{\thesection.\arabic{equation}}   
\setcounter{equation}{7}                                   
 At the points $\vec{x}_{1,2}$ the potentials 
  $ \widetilde{A}\ ^{a}_{i}$ coincide with the expressions
given in (3.7a--b), because the regularization of
$\partial_{i}\phi$ leads to the vanishing of the self
interaction terms in (3.7c). Furthermore the nonlinear term in the  definition
  of
$ \widetilde{F}\ ^{a}_{ij}$  (see in (3.6)) vanishes at
$\vec{x}_{1,2}$, what completes the proof.

The relation  (3.5) will not be discussed further as $A^{a}_{0}$
is not needed in the following.

By the same procedure as described in Section 2 we may now
derive the expressions for the two--body Hamiltonian $H$ and the
canonical momentum for the relative particle motion $\vec{p}$. We obtain
\begin{equation}
H = \xi_{i} \, \xi_{i}\, ,
\end{equation}
and
\begin{equation}
\xi_{i} =  p_{i} - {\lambda \over 4\pi} \,
\partial_{i}(\vec{x})
\left(
H - {2\over \lambda\kappa} \, Q^{a}_{1}Q^{a}_{2}
\right)\, .
\end{equation}
Note that (3.8) and (3.9) are the same as (2.9) and (2.10)
respectively with the electric charges $e_{\alpha}$ replaced by their
 $SU(2)$ counterparts.

 Therefore, we obtain by squaring (3.9) again
 \begin{equation}
H = p^{2} - {l^{2}\over r^{2}} +   {\overline{l}^{2}\over
r^{2}}\, ,
\end{equation}
with
\begin{equation}
l = \overline{l} + {\lambda\over 4\pi}
\left( H -  {2\over   \lambda \kappa}\, Q^{a}_{1}\, Q^{a}_{2}\right)\, ,
\end{equation}
where $l$ and $\overline{l}$ are defined as before.

Finally, we have to show that $ Q^{a}_{1}Q^{a}_{2}$ is a
conserved quantity,{\it  i.e.}
\begin{equation}
{d\over dt} \left(  Q^{a}_{1}Q^{a}_{2}\right)
= 0\, .
\end{equation}
In order to prove this  statement we start with the non--Abelian 
 counterpart to
(2.14) given by
\begin{equation}
L = L_{0} -
{ Q^{a}_{1}Q^{a}_{2}\over 2\pi\kappa} \, {d\over dt}
\, \phi(\vec{x}) + L_{\rm {SU(2)}}\, .
\end{equation}
Then the Euler--Lagrange equations for $ Q^{a}_{\alpha}$ are the
 Wong--equations [10,11] which take the form
 \begin{equation}
 \dot{Q}^{a}_{1} - {\dot{\phi}\over 2\pi\kappa} \,
 \epsilon_{abc} \,  Q^{b}_{2}Q^{c}_{1} = 0\, .
 \end{equation}
 By exchanging the particle indices 1 and 2   we get
  \begin{equation}
 \dot{Q}^{a}_{2} + {\dot{\phi}\over 2\pi\kappa} \,
 \epsilon_{abc} \,  Q^{b}_{2}Q^{c}_{1} = 0\, .
 \end{equation}
Note that (3.14-15) imply that the lengths of $Q_{\alpha}$ are conserved.
Thus we  conclude that
  \begin{equation}
 \dot{Q}^{a}_{1} +  \dot{Q}^{a}_{2}  = 0\, ,
 \end{equation}
which, due to (1.8), leads  to the desired result (3.12).
With (3.12) we  conclude  from (3.13) that the formulae (2.15--17) 
 hold as  in   the non--Abelian case.

 \section{The Quantum--Mechanical Two--Body Problem on a Plane}

Let us start with the observation that we have, in both 
the Abelian and  non--Abelian cases, the same structure of the
classical two--body Hamiltonian

\begin{equation}
H = p^{2} - {l^{2}\over r^{2}} +   {\overline{l}^{2}\over
r^{2}}\, ,
\end{equation}
with

\begin{equation}
l = \overline{l} + {\lambda\over 4\pi}
\left( H -  {\Omega \over \lambda \kappa}\right)\, ,
\end{equation}
with $\Omega$, a function of the particle charges, given by

\begin{equation}
\Omega = \left\{ \begin{array}{ll}
2e_{1}\, e_{2}\qquad & \mbox{Abelian case} \cr
2Q^{a}_{1}\, Q^{a}_{2}\qquad & \mbox{non--Abelian case}
\end{array}
\right. \, .
\end{equation}
Without the gauge fields we have $\Omega=0$. Therefore, in
quantizing (4.1--2) we can follow the techniques presented in [1,12]. 
 However, 
 we should keep in mind that the quantum theory requires a
quantized coupling $\kappa $  in the non--Abelian case  [8,17]
\begin{equation}
4\pi \kappa \in \Bbb{Z} \, .
\end{equation}
Moreover, we have to properly take into account, the quantum nature 
of the operator $\hat \Omega$. 
\textheight 8.7in \textwidth 6in
We quantize the problem  by considering
 a Schr\"odinger-like equation 
\begin{equation}
i\hbar {\partial \psi(\vec{x},t) \over 
\partial t} =    \hat{H} \, \psi (\vec{x},t) =
\left[
\hat{\vec{p}}\, \sp2- {l^2  \over r^2} +
{\overline{l}\sp2\over  r^2 } \right]
\Psi(\vec{x},t)
\label{eq6}
\end{equation}
in which the operators $\hat H$ and $\hat{\vec{p}}$ 
  are defined  by  the usual
quantization rules 
\begin{equation}
\hat H\,= \, i\hbar {\partial\over \partial t},\qquad
 \hat{p}_{i}\,= \,
{\hbar\over i}{\partial\over \partial x_i} \, .
\end{equation}

Note that, in the non-Abelian case, the eigenvalues 
of $\hat \Omega$  (see (1.8)) are determined by
\begin{equation}
\hat\Omega\vert j_{12},j_1,j_2>=\Omega\vert j_{12},j_1,j_2>=
 {1\over2}\left( j_{12}(j_{12}+1
)-j_1(j_1+1)-j_2(j_2+1)\right)\vert j_{12},j_1,j_2>
\end{equation} 
and so the wavefunction $\Psi(\vec{x},t)$ depends also on the eigenvalues
$j_1$, $j_2$ and $j_{12}$.
More explicitly, the wave function in (4.5), for two
non-Abelian CS particles with definite quantized isospins
$j_1$ and $j_2$ describes a multiplet
of wave functions with $n$ components, where
$n=\vert j_1+j_2\vert -\vert j_1-j_2\vert +1$.  This
multiplet structure and the eigenvalues of $\hat \Omega$
given by (4.7) will be implicitly assumed
in all formulae that follow.

For the stationary case, {\it i.e.}
 when $\Psi(\vec{x},t)=\Psi_E(\vec{x})e\sp{{iEt\over \hbar}}$
we can use the angular-momentum basis and put
\begin{equation}
\Psi_{E,m}\,=\,f_{E,m}(r ) \, e\sp{im\varphi}
\end{equation}
where $m$ is an integer, and find that $f_{E,m}$ satisfies a nonstandard
 time independent  Schr\"odinger equation
\begin{equation}
\left[-\hbar\sp2\left(\partial_r\sp2 
+ {1\over r}\partial_r 
-{\bar{m}\sp2\over r\sp2}\right)-E\right]f_{E,m}(r) =0,
\label{sch}
\end{equation}
where, in consistency  with (4.2),  we have defined
\begin{equation}
\hbar \bar{m} :\,=\,\hbar m-{\lambda\over 4\pi}\left( E-{\Omega\over
 \lambda\kappa}\right)
\label{bar}
\end{equation}
{\it i.e.} $\hbar \bar{m}$ is an eigenvalue of ${\overline{l}}$ and
$\Omega$ denotes the eigenvalue of $\hat \Omega$. 

We see that our equation (4.9) is the same as the equation 
(7.6) of [12] with an important difference due to $\Omega$.
The existence of $\Omega$ leads not only to  the redefinition
of $\bar{m}$ but, in the non-Abelian case, also to the splitting
 of the energy levels  as for any integer
or half-interger values of $j_1$ and $j_2$ there are several
values of $j_{12}$ which satisfy $j_{12}\in (\vert j_1-j_2\vert,\vert
 j_1+j_2\vert)$. To determine the energy 
levels,  we can, however, follow the procedure used in [12]. 
Thus, in particular, if we focus our attention
on the interior solutions ($r<r_0$) we find that they are given by
\begin{equation}
f_{E,m}(r) = 
J_{ \overline{m}} \left( {\sqrt{E} \over \hbar} \, r \right)
\, ,
\end{equation} 
(restricting our attention to the more interesting case of $\lambda\bar{l}<0$)

The energy levels are then given by the eigenvalues of $H$, which
 are determined
by the boundary condition corresponding to the
requirement that the wavefunction vanishes at $r=r_0$, 
and are given by:

\begin{equation}
J_{\bar{m}}\left[{\sqrt{E}\over \hbar}\left({\hbar\vert
 \lambda\bar{m}\vert\over 2\pi}\right)\sp{1\over 2}\right]\,=\,0
\label{result}
\end{equation}
with $\bar{m}$ given by (\ref{bar}).

Let us look  at the case $\bar{m}> 0$, $\lambda<0$.
Then it is convenient to define
\begin{equation}
\epsilon\,=\,{\vert \lambda\vert E\over 2\pi \hbar}
\end{equation}
 so that (\ref{result}) takes the form
\begin{equation}
J_{\bar{m}}(\bar{m}\sp{1\over 2}\epsilon\sp{1\over2})\,=\,0.
\end{equation}

The Bessel function  $J_{\bar{m}}$, for fixed $\bar{m}>0$, has
 an infinite number of positive
zeroes which, in what follows, we denote by $y_n(\bar{m})$, $n=1,2..$ Thus
 we see that due to (\ref{bar}), the eigenvalues $\epsilon_n(m)$  are the
 positive fixed points of the equation
\begin{equation}
\epsilon\,=\,f_n(\bar{m})\,=\,f_n\left(m+{1\over 2}
\epsilon+{\Omega\over 4\pi\hbar\kappa}\right),
\label{neweq}
\end{equation}
where  
\begin{equation}
f_n(\bar{m})\,=\,{1\over \bar{m}}y_n\sp2(\bar{m}).
\end{equation}

The existence of positive fixed points $\epsilon$ of (\ref{neweq}) 
was discussed in great detail in [12], both by using various asymptotic 
formulae for the zeroes of Bessel's functions and also by solving 
(\ref{neweq}) numerically. As the present case differs from the case
without the gauge functions 
by the redefinition of $\bar{m}$, below we present the figure from [12],
but this time with the interpretation that the horizontal axis denotes not $m$
but $\left(m+{\Omega\over 4\pi\hbar\lambda\kappa}\right)$.

 The plot looks like several curves; the lowest values correspond
to the first zeroes ({\it ie} $n=1$), the next ones to second zeroes {\it ie}
 $n=2$ {\it etc}. The points lie so close that the figure may appear as a set of lines while,
in reality, we have here sets of points. The points appear to be 
(almost) equally spaced on each ``curve" - this is due to the 
approximate linearity of the positions of zeros of Bessel functions
as a function of $\bar{m}$. To check our values of energies we had also
solved (\ref{neweq}) differently; we approximated the positions
of the zeroes of the Bessel functions by a linear function and solved
the resultant equations for $\epsilon$. The obtained results were very similar
to those of our figure thus giving us confidence in our results.

Our results show that, for each value of $m$
and so for each value of $\tilde m=\left(m+{\Omega\over 4\pi\hbar\lambda\kappa}\right)$,
 there is a whole tower of
values of $\epsilon$ corresponding to different zeroes of the Bessel functions.
In addition, in the non-Abelian case, there is a further splitting of energy
levels due to the different values of $j_{12}$ in $\Omega$.
The values of $\epsilon$ increase, approximately linearly, as we take
higher zeros ({\it ie} $y_n$ for larger $n$). The dependence on $m$ is
 only slightly more complicated;
for each order of the zero there is a value of $m$ for which the energy
is minimal and as we move away from this value the energy grows, approximately,
linearly. As $n$ increases the minimal values of $m$ increase, again, approximately linearly.

\begin{figure}[h]
\unitlength1cm
\hfil\begin{picture}(12,12)
\epsfxsize=11cm
\epsffile{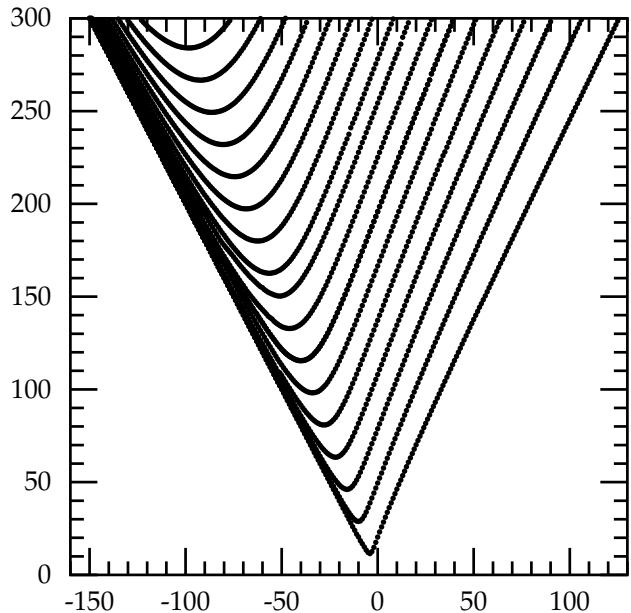}
\end{picture}
\caption{Energy as a function of $\tilde m=\left(m+{\Omega\over 4\pi\hbar\lambda\kappa}\right)$}
\end{figure}

 Note  that for $\bar{m}<0$
and $\lambda>0$ the corresponding energy levels are obtained by changing the sign of $m$ and $\Omega$.

We summarise our results by noting that in the interior region $r<r_0$,
 where classical solutions are only possible for a finite time interval, we
 can find quantum solutions which correspond to discrete bound states 
determined by the boundary condition at $r=r_0$.  This shows that as in 
the case discussed in [12]  this boundary
condition defines a sort of planar geometric ``bag" for the quantum state.

The discussion of the exterior solutions is again similar to what we presented
in [12]. The system has no bound states and the
scattering solutions are given by 
a superposition of Bessel functions of the first and second kind

\begin{equation}
f_{E,m}(r )\,=\,A_m(E)\,J_{\bar{m}}\left({\sqrt{E}\over \hbar}r\right)\,+\,
B_m(E)\,Y_{\bar{m}}\left({\sqrt{E}\over \hbar}r\right)
\end{equation}
with the ratio ${A_m\over B_m}$ determined by the boundary condition 
of the wavefunction vanishing at $r=r_0$.
Clearly the solutions describe a scattering
  on an obstruction of radius $r_0$, which is
dynamically determined.

\section{Final Remarks}
In D=2+1 dimensions one can consider four basic actions describing
gravitational and gauge degrees of freedom\footnote{One can consider also
models with linear combinations of Maxwell and Chern--Simons terms
in the gauge sector as well as both the Einstein and 
translational Chern--Simons
 terms in the gravity sector (cp.[17]). 
The (2+1) dimensional gravity with the Einstein term supplemented by the 
translational Chern--Simons term was named ``vector Chern--Simons gravity''
in [18]}.
\begin{itemize}
\item{} Einstein action, linear in Riemann curvature with Maxwell or Yang-Mills gauge fields
\item{} Einstein action with (Abelian or non-Abelian) Chern--Simons gauge fields
\item{} Translational Chern--Simons gravity action with Maxwell or Yang Mills gauge fields
\item{} Translational Chern--Simons gravity with Chern--Simons gauge fields.
\end{itemize}

In this paper we have studied the last case in this list and considered
 the coupling
to $D=2+1$ nonrelativistic particles. We have shown that in the interacting
$D=2+1$ Chern--Simons theories, with sources, the field equations take the form
of field-current identities. This has allowed us to eliminate the field
degrees of freedom and to obtain, without any approximation, the 
planar two-body interaction.

In the non-Abelian case, following [11], we have considered the motion of
 particles in a two-dimensional space extended by internal coordinates,
 in accordance 
with the Kaluza-Klein approach to internal symmetries.

Our basic result is a quantum-mechanical solution 
of the 2-body problem, describing dynamically confined particles,
 with the energy-dependent potential
generated by the {\bf double} (gravitational and gauge)
 Chern--Simons couplings.
In our previous papers [1,12] we showed that the (single) gravitational
Chern--Simons coupling in $D=2$ dimensions
\begin{itemize}
\item{} leads to planar confinement
\item{} implies the noninteger values of the quantum number $\bar m_0$
\end{itemize}
\begin{equation}
\bar m_0\,=\,m\,-\,{\lambda E\over 4\pi \hbar},
\end{equation}
describing the continuous values of the Abelian spin in $D=2+1$ dimensions. 
We see from (5.1) that the gravitational Chern--Simons coupling leads to the
anyonic behaviour of massive point particles.\footnote{Such an 
observation was made earlier in [19]}

The effect of adding the Chern--Simons gauge interaction in both the Abelian
and non-Abelian cases reduces to the additional shift of the continuous
 Abelian spin value
\begin{equation}
\bar m_0\,\rightarrow\,\bar m\,=\,\bar m_0\,+\, {\Omega\over 4\pi\kappa\hbar}.
\end{equation}

Thus we see that in the Abelian case we have anyonic values of the angular
momentum, shifted by a term proportional to the product of Abelian 
gauge charges. In the non-Abelian case the shift is given by the 
eigenvalues of the operator ${\hat \Omega\over2}$ (1.8), described
explicitly by (4.7) {\it ie} it is matrix valued. We see from our results
that the planar confinement remains valid as in the purely
gravitational case. In the non-Abelian case we have a new effect - the
splitting of energy levels.

\subsection*{Acknowledgment}
The work reported in this paper was completed when two of us (JL and WJZ)
were visiting, respectively, the University of Valencia, Spain and MIT, USA.
They would like to thank their hosts for their hospitality and for
 financial support making their visits possible, respectively,
JL - Generalitat Valenciana and WJZ - Center for Theoretical Physics, MIT.

\textheight 8.8in \textwidth 6in

\end{document}